\documentclass[12pt]{article}
\usepackage {axodraw}

\catcode`@=12
\begin{document}
\thispagestyle{empty}
\begin{flushright} 
UCRHEP-T356\\ 
June 2003\
\end{flushright}
\vspace{0.5in}
\begin{center}
{\LARGE	\bf New ``Square Root'' Model of\\ Lepton Family Cyclic Symmetry\\} 
\vspace{1.5in}
{\bf Ernest Ma$^a$ and G. Rajasekaran$^b$\\}
\vspace{0.2in}
{\sl $^a$ Physics Department, University of California, Riverside, 
California 92521, USA\\}
\vspace{0.1in}
{\sl $^b$ Institute of Mathematical Sciences, Chennai (Madras) 600113, India\\}
\vspace{1.5in}
\end{center}
\begin{abstract}\
Following the newly formulated notion of form invariance of the neutrino 
mass matrix, a complete model of leptons is constructed.  It is based on a 
specific unitary $3 \times 3$ matrix $U$ in family space, such that $U^2$ is 
the simple discrete symmetry $\nu_e \to -\nu_e$, $\nu_\mu \leftrightarrow 
\nu_\tau$.  Thus $U$ also generates the cyclic group $Z_4$. The charged-lepton 
mass matrix is nearly diagonal while the neutrino mass matrix is of the form 
suitable for explaining maximal (large) mixing in atmospheric (solar) 
neutrino oscillations in the context of three nearly degenerate 
neutrino masses.  Observable lepton flavor violation is predicted.
Quarks may be treated in the same way as the charged leptons.
\end{abstract}
\newpage
\baselineskip 24pt

To understand the form of the neutrino mass matrix ${\cal M}_\nu$, a new 
idea has recently been proposed \cite{ma03}.  It is postulated that there 
is a specific $3 \times 3$ unitary matrix $U$ with $U^{\bar n} = 1$ such that
\begin{equation}
U {\cal M}_\nu U^T = {\cal M}_\nu,
\end{equation}
and for some $N < \bar n$, the matrix $U^N$ represents a well-defined 
discrete symmetry in the $\nu_{e,\mu,\tau}$ basis, for which the 
charged-lepton mass matrix ${\cal M}_l$ is diagonal.  However, since each 
neutrino belongs to an $SU(2)_L \times U(1)_Y$ doublet, the corresponding 
left-handed charged leptons ($e, \mu, \tau$) must also transform under $U$.  
A complete theory must then reconcile the apparent contradictory requirement 
that ${\cal M}_\nu$ satisfies Eq.~(1), but ${\cal M}_l$ does not.  The 
resolution of this conundrum is in the soft and spontaneous breaking of 
the symmetry supported by $U$, as was done for example in the $A_4$ model 
\cite{mara01,bmv} of degenerate neutrino masses.  In this paper we show how 
this may be achieved with the specific unitary matrix
\begin{equation}
U = \pmatrix {0 & i/\sqrt 2 & -i/\sqrt 2 \cr i/\sqrt 2 & 1/2 & 1/2 \cr 
-i/\sqrt 2 & 1/2 & 1/2},
\end{equation}
where
\begin{equation}
U^2 = \pmatrix {-1 & 0 & 0 \cr 0 & 0 & 1 \cr 0 & 1 & 0},
\end{equation}
i.e. the simple discrete symmetry
\begin{equation}
\nu_e \to -\nu_e, ~~~ \nu_\mu \leftrightarrow \nu_\tau.
\end{equation}
The matrix $U$ of Eq.~(2) is thus a ``square root'' of this discrete 
symmetry.  We see immediately also that $U^4 = 1$; hence our model is a 
specific realization of the cyclic group $Z_4$ as a family symmetry.

The most general Majorana mass matrix ${\cal M}_\nu$ is of the form
\begin{equation}
{\cal M}_\nu = \pmatrix {A & D & E \cr D & B & F \cr E & F & C}.
\end{equation}
Using Eqs.~(1) and (2), we find that it becomes
\begin{equation}
{\cal M}_\nu = \pmatrix {A & 0 & 0 \cr 0 & B & A+B \cr 0 & A+B & B}.
\end{equation}
This mass matrix has eigenvalues $A$, $-A$, and $A+2B$, corresponding to the 
eigenstates $\nu_e$, $(\nu_\mu-\nu_\tau)/\sqrt 2$, and $(\nu_\mu+\nu_\tau)
/\sqrt 2$.  We see immediately that $\nu_\mu - \nu_\tau$ mixing is maximal 
with
\begin{equation}
\Delta m^2_{atm} = (A+2B)^2 - A^2 = 4B(A+B),
\end{equation}
which is suitable for explaining atmospheric neutrino oscillations \cite{atm}. 
If $A << B$, we have the hierachical structure of neutrino masses and 
$B = \sqrt {\Delta m^2_{atm}}/2 \simeq 0.025$ eV.  On the other hand, if 
$B << A$, we have the more interesting scenario of three nearly degenerate 
neutrino masses, with the prediction that $A$ is large enough to be measured 
by neutrinoless double beta decay.  As for solar neutrino oscillations 
\cite{solar}, we can obtain the large-mixing-angle solution by invoking 
flavor-changing radiative corrections \cite{ma03,bmv}.  Details will be 
presented in a later paragraph.

The leptonic Yukawa couplings of our model is given by
\begin{eqnarray}
{\cal L}_Y &=& h_{ij} [\xi^0 \nu_i \nu_j - \xi^+ (\nu_i l_j + l_i \nu_j)/
\sqrt 2 + \xi^{++} l_i l_j] \nonumber \\ 
&+& f_{ij}^k (l_i \phi_j^0 - \nu_i \phi_j^-) l_k^c + H.c.,
\end{eqnarray}
where we have adopted the convention that all fermion fields are left-handed, 
with their right-handed counterparts denoted by the corresponding 
(left-handed) charge-conjugate fields.  We have also extended the Higgs 
sector of the Standard Model of particle interactions to include three 
doublets $(\phi_j^0, \phi_j^-)$ and one very heavy triplet $(\xi^{++}, \xi^+, 
\xi^0)$.  The electroweak $SU(2)_L \times U(1)_Y$ gauge symmetry will be 
spontaneously broken mainly by the vacuum expectation value of one particular 
Higgs doublet.  As a result, the vacuum expectation values of the other two 
Higgs doublets can be naturally small \cite{ma01} and that of the triplet 
even smaller \cite{masa98}.

We now assume that ${\cal L}_Y$ is invariant under the transformation
\begin{eqnarray}
&& (\nu,l)_i \to U_{ij} (\nu,l)_j, ~~~ l^c_k \to l^c_k, \\ 
&& (\phi^0,\phi^-)_i \to U_{ij} (\phi^0,\phi^-)_j, ~~~ (\xi^{++},\xi^+,\xi^0) 
\to (\xi^{++},\xi^+,\xi^0).
\end{eqnarray}
This means
\begin{equation}
U^T h \,U = h, ~~~ U^T f^k \,U = f^k,
\end{equation}
resulting in
\begin{equation} 
h = \pmatrix {a & 0 & 0 \cr 0 & b & a+b \cr 0 & a+b & b}, ~~~ 
f^k = \pmatrix {a_k & d_k & -d_k \cr -d_k & b_k & a_k+b_k \cr d_k & a_k+b_k 
& b_k}.
\end{equation}
Note that $h$ has no $d$ terms because it has to be symmetric.

The neutrino mass matrix ${\cal M}_\nu$ is then given by
\begin{equation}
{\cal M}_\nu = 2 h \langle \xi^0 \rangle,
\end{equation}
whereas the charged-lepton mass matrix ${\cal M}_l$ linking $l_i$ to $l^c_k$ 
is
\begin{equation}
\pmatrix {a_1 v_1 + d_1 (v_2-v_3) & a_2 v_1 + d_2 (v_2-v_3) & 
a_3 v_1 + d_3 (v_2-v_3) \cr -d_1 v_1 + b_1 (v_2+v_3) + a_1 v_3 & -d_2 v_1 
+ b_2 (v_2+v_3) + a_2 v_3 & -d_3 v_1 + b_3 (v_2+v_3) + a_3 v_3 \cr 
d_1 v_1 + a_1 v_2 + b_1 (v_2+v_3) & d_2 v_1 + a_2 v_2 + b_2 (v_2+v_3) & 
d_3 v_1 + a_3 v_2 + b_3 (v_2+v_3)},
\end{equation}
where $v_i \equiv \langle \phi_i^0 \rangle$.  Assume $d_k,b_k << a_k$ and 
$v_{1,3} << v_2$, then all elements in the first, second, and third rows 
are of order $d v_2 + a v_1$, $b v_2 + a v_3$, and $a v_2$ respectively. 
It is clear that they may be chosen to be of order $m_e$, $m_\mu$, and 
$m_\tau$, in which case ${\cal M}_l$ will become nearly diagonal by simply 
redefining the $l^c_k$ basis.  The mixing matrix $V_L$ in the $l_i$ basis 
(such that $V_L {\cal M}_l {\cal M}_l^\dagger V_L^\dagger$ is diagonal) 
will be very close to the identity matrix with off-diagonal terms of order 
$m_e/m_\mu$, $m_e/m_\tau$, and $m_\mu/m_\tau$.  This construction allows 
us then to consider ${\cal M}_\nu$ to be in the basis $(\nu_e,\nu_\mu,
\nu_\tau)$ to a very good approximation.

We now understand why it is sensible \cite{ma03} to consider Eq.~(1) as a 
condition on ${\cal M}_\nu$.  The key lies in the fact that ${\cal M}_\nu$ 
comes from neutrino couplings to a single field $\xi^0$ which is invariant 
under $U$, whereas ${\cal M}_l$ comes from couplings to $\phi^0_{1,2,3}$ 
which are not.  Let the scalar trilinear coupling of 
$\xi$ to $\Phi_2 \Phi_2$ be $\mu$, then \cite{masa98}
\begin{equation}
\langle \xi^0 \rangle \simeq {-\mu v_2^2 \over M_\xi^2},
\end{equation}
which shows clearly that neutrino masses may be of order 1 eV or less if 
$M_\xi^2/\mu \sim 10^{13}$ GeV.  Similarly, $v_{1,3}$ can be small compared 
to $v_2$ if $M_2^2 < 0$ but the $M_{1,3}^2$ terms in the Higgs potential 
are large and positive, in which case \cite{ma01}
\begin{equation}
v_1 \simeq {-\mu_{12}^2 v_2 \over M_1^2}, ~~~ v_3 \simeq {-\mu_{23}^2 v_2 
\over M_3^2},
\end{equation}
where $\mu_{12}^2$ and $\mu_{23}^2$ are the coefficients of the 
$\Phi_1^\dagger \Phi_2$ and $\Phi_2^\dagger \Phi_3$ terms respectively.

Going back to ${\cal M}_\nu$ of Eq.~(6), we now consider how solar neutrino 
oscillations may arise in the $2 \times 2$ submatrix spanning $\nu_e$ and 
$(\nu_\mu - \nu_\tau)/\sqrt 2$, i.e.
\begin{equation}
{\cal M} = \pmatrix {A & 0 \cr 0 & -A}.
\end{equation}
Consider the most general radiative corrections to ${\cal M}$, i.e.
\begin{equation}
R = \pmatrix {r_{11} & r_{12} \cr r_{12}^* & r_{22}},
\end{equation}
then ${\cal M}$ becomes
\begin{equation}
(1+R) {\cal M} (1+R^T) \simeq A \pmatrix {1+2r_{11} & r_{12}^* - r_{12} \cr 
r_{12}^* - r_{12} & -1-2r_{22}}.
\end{equation}
In terms of the full radiative correction matrix \cite{bmv},
\begin{eqnarray}
&& r_{11} = r_{ee}, ~~~ r_{22} = {1 \over 2} (r_{\mu \mu} + r_{\tau \tau}) - 
Re (r_{\mu \tau}), \\ && r_{12}^* - r_{12} = -2i Im (r_{12}) = -i \sqrt 2 
[Im (r_{e \mu}) - Im (r_{e \tau})]. 
\end{eqnarray}
Thus the radiatively corrected ${\cal M}$ has eigenvalues
\begin{equation}
m_{1,2} = A \left( 1+r_{11}+r_{22} \mp \sqrt {(r_{11}-r_{22})^2 + 4[Im 
(r_{12})]^2}
\right)
\end{equation}
corresponding to the eigenvectors $\nu_e \cos \theta - i (\nu_\mu - \nu_\tau)
\sin \theta /\sqrt 2$ and $\nu_e \cos \theta + i (\nu_\mu - \nu_\tau) 
\cos \theta /\sqrt 2$ respectively, where
\begin{equation}
\tan \theta = {r_{11} - r_{22} + \sqrt {(r_{11}-r_{22})^2 + 4[Im (r_{12})]^2} 
\over 2 |Im (r_{12})|},
\end{equation}
with $r_{22} - r_{11} > 0$.  Since $\tan^2 \theta \simeq 0.46$ is desirable 
\cite{msv} for understanding solar neutrino oscillations \cite{solar}, 
flavor changing ($r_{12} \neq 0$) and flavor nonuniversal ($r_{11} \neq 
r_{22}$) interactions are required.  Specific examples have already been 
proposed \cite{ma03,bmv}.  As for $\Delta m^2_{sol}$, it is given here by
\begin{equation}
\Delta m^2_{sol} = m_2^2 - m_1^2 \simeq 4A^2 \sqrt {(r_{11}-r_{22})^2 + 
4[Im (r_{12})]^2}.
\end{equation}
For $|r| \sim 10^{-3}$ which is a typical size for radiative corrections and 
$\Delta m^2_{sol} \simeq 6.9 \times 10^{-5}$ eV$^2$ \cite{msv}, $|A| \sim 
0.13$ eV is then obtained.  [The recent WMAP result implies \cite{wmap} an 
upper bound of 0.23 eV on $|A|$ from cosmological considerations.]  Given 
that $|A|$ is also the effective neutrino mass measured in neutrinoless 
double beta decay with a present upper bound of about 0.4 eV, this is an 
encouraging prediction for future 
experiments \cite{klapdor}.  If $A \simeq 0.13$ eV, then using Eq.~(7), we 
find $B \simeq 0.0048$ eV and $B/A \simeq 0.037$ which is of the same order 
as $m_\mu/m_\tau \simeq 0.059$, as suggested by ${\cal M}_l$ of Eq.~(14).

With flavor changing radiative corrections, the $U_{e3}$ entry of the neutrino 
mixing matrix becomes nonzero. It is given here by
\begin{equation}
U_{e3} \simeq -{[Re (r_{e \mu}) + Re (r_{e \tau})] A \over \sqrt 2 B},
\end{equation}
which may be as large as the experimental upper bound \cite{react} of 0.16. 
However, it is real so that there is no $CP$ violation.  This is in contrast 
to the case of the $A_4$ model \cite{bmv}, where $U_{e3}$ has to be purely 
imaginary \cite{mgl}.

Going back to the Yukawa couplings of the leptons to the 3 Higgs doublets 
given by Eq.~(12) and assuming the hierarchy $d_k << b_k << a_k$ and taking 
the limit that only $v_2$ is nonzero, we have ${\cal M}_l$ of Eq.~(14) 
simply given by
\begin{equation}
{\cal M}_l \simeq v_2 \pmatrix {d_1 & d_2 & d_3 \cr b_1 & b_2 & b_3 \cr 
a_1 & a_2 & a_3},
\end{equation}
whereas $\Phi_1$ and $\Phi_3$ couple to $l_i l^c_j$ according to
\begin{equation}
\pmatrix {a_1 & a_2 & a_3 \cr -d_1 & -d_2 & -d_3 \cr d_1 & d_2 & d_3}, ~~~ 
\pmatrix {-d_1 & -d_2 & -d_3 \cr a_1 & a_2 & a_3 \cr b_1 & b_2 & b_3},
\end{equation}
respectively.  After rotating ${\cal M}_l$ of Eq.~(26) in the $l^c_j$ basis 
to define the state corresponding to $\tau$, we see immediately from Eq.~(27) 
that the dominant coupling of $\Phi_1$ is $(m_\tau/v_2) e \tau^c$ and that 
of $\Phi_3$ is $(m_\tau/v_2) \mu \tau^c$.  Other couplings are at most of  
order $m_\mu/v_2$ in this model, and some are only of order $m_e/v_2$.  We 
thus have a natural understanding of the smallness of flavor changing decays 
in the leptonic sector, even though they should be present and are 
potentially observable.

Using Eq.~(27), we see that the decays $\tau^- \to e^- e^+ e^-$ and $\tau^- 
\to e^- e^+ \mu^-$ may proceed through $\phi_1^0$ exchange with coupling 
strengths of order $m_\mu m_\tau/v_2^2 \simeq (g^2/2) (m_\mu m_\tau/M_W^2)$, 
whereas the decays $\tau^- \to \mu^- \mu^+ \mu^-$ and $\tau^- \to \mu^- \mu^+ 
e^-$ may proceed through $\phi_3^0$ exchange also with coupling strengths of 
the same order.  We estimate the order of magnitude of these branching 
fractions to be
\begin{equation}
B \sim \left( {m_\mu^2 m_\tau^2 \over M_{1,3}^4} \right) B(\tau \to \mu \nu 
\nu) \simeq 6.1 \times 10^{-11} \left( {100~{\rm GeV} \over M_{1,3}} \right)^4,
\end{equation}
which is well below the present experimental upper bound of the order 
$10^{-6}$ for all such rare decays \cite{pdg}.

The decay $\mu^- \to e^- e^+ e^-$ may also proceed through $\phi_1^0$ with 
a coupling strength of order $m_\mu^2/v_2^2$.  Thus
\begin{equation}
B(\mu \to e e e) \sim {m_\mu^4 \over M_1^4} \simeq 1.2 \times 10^{-12} \left( 
{100~{\rm GeV} \over M_1} \right)^4,
\end{equation}
which is at the level of the present experimental upper bound of $1.0 \times 
10^{-12}$.  The decay $\mu \to e \gamma$ may also proceed through $\phi_3^0$ 
exchange (provided that $Re \phi_3^0$ and $Im \phi_3^0$ have different masses) 
with a coupling of order $m_\mu m_\tau/v_2^2$.  Its branching fraction is 
given by \cite{mara01}
\begin{equation}
B(\mu \to e \gamma) \sim {3 \alpha \over 8 \pi} {m_\tau^4 \over M_{eff}^4},
\end{equation}
where
\begin{equation}
{1 \over M_{eff}^2} = {1 \over M_{3R}^2} \left( \ln {M_{3R}^2 \over m_\tau^2} 
- {3 \over 2} \right) - {1 \over M_{3I}^2} \left( \ln {M_{3I}^2 \over 
m_\tau^2} - {3 \over 2} \right).
\end{equation}
Using the experimental upper bound \cite{meg} of $1.2 \times 10^{-11}$, we 
find $M_{eff} > 164$ GeV.

In the quark sector, if we use the same 3 Higgs doublets for the corresponding 
Yukawa couplings, the resulting $up$ and $down$ mass matrices will be of the 
same form as Eq.~(14).  Because the quark masses are hierarchical in each 
sector, we will also have nearly diagonal mixing matrices as in the case of 
the charged leptons.  This provides a qualitative understanding in our model 
of why the charged-current mixing matrix linking $up$ quarks to $down$ quarks 
has small off-diagonal entries.

Once $\phi_1^0$ or $\phi_3^0$ is produced, its dominant decay will be to 
$\tau^\pm e^\mp$ or $\tau^\pm \mu^\mp$ if each couples only to leptons. 
If they also couple to quarks (and are sufficiently heavy), then the dominant 
decay products will be $t \bar u$ or $t \bar c$ together with their 
conjugates.  As for $\phi_2^0$, it will behave very much as the single 
Higgs doublet of the Standard Model, with mostly diagonal couplings to 
fermions.

In conclusion we have constructed a complete theory of leptons where the 
neutrino mass matrix ${\cal M}_\nu$ may be derived from the requirement 
that $U {\cal M}_\nu U^T = {\cal M}_\nu$, where $U$ is a specific $3 \times 3$ 
unitary matrix in family space such that $U^2$ is the simple discrete 
symmetry $\nu_e \to -\nu_e$, $\nu_\mu \leftrightarrow \nu_\tau$.  We obtain 
three nearly degenerate neutrino masses with maximal mixing for atmospheric 
neutrino oscillations.  Solar neutrino oscillations are induced by 
flavor-changing radiative corrections.  As a result, neutrinoless double 
beta decay is predicted to occur at the 0.1 eV range.  There are also three 
Higgs doublets in this model, two of which have dominant flavor-changing 
couplings proportional to $m_\tau$ and may be easily observed at future 
colliders.

This work was supported in part by the U.~S.~Department of Energy
under Grant No.~DE-FG03-94ER40837.  G.R. also thanks the UCR Physics 
Department for hospitality and acknowledges DAE-BRNS (India) for support. 
We also thank the Kavli Institute for Theoretical Physics, U. C. Santa 
Barbara, for hospitality during its 2003 Neutrino Workshop where this work 
was initiated.

\bibliographystyle{unsrt}

\end{document}